\documentclass[twocolumn]{aastex62}
\usepackage{apjfonts}
\usepackage{amsmath}

\newcommand{\Hi}{\hbox{H{\sc i}}}
\newcommand{\hii}{\textup{H}\,\textsc{ii}}

\begin{document}

\title{The H$\alpha$ concentration of local star-forming galaxies: implications for galaxy structure formation}
\author{Zhizheng Pan}
\email{panzz@pmo.ac.cn}
\affiliation{Purple Mountain Observatory, Chinese Academy of Sciences, 10 Yuan Hua Road, Nanjing, Jiangsu 210033, China}
\affiliation{School of Astronomy and Space Sciences, University of Science and Technology of China, Hefei, 230026, China
}

\author{Xianzhong Zheng}
\affiliation{Tsung-Dao Lee Institute and Key Laboratory for Particle Physics, Astrophysics and Cosmology, Ministry of Education, Shanghai Jiao Tong University, Shanghai, 201210, China}

\author{Xu Kong}
\affiliation{School of Astronomy and Space Sciences, University of Science and Technology of China, Hefei, 230026, China
}
\affiliation{Department of Astronomy, \\
University of Science and Technology of China, Hefei, Anhui 230026, China}

\begin{abstract}
In this work, we present a study on the H$\alpha$ emission line flux concentration of 3098 low-redshift star-forming galaxies (SFGs) using the MaNGA data available in the Data Release 17 from the Sloan Digital Sky Survey. We define the H$\alpha$ flux concentration index ($C_{\rm H\alpha}$) as $C_{\rm H\alpha}=F_{\rm H\alpha,0.8~R_e}/F_{\rm H\alpha,1.5~R_e}$, where $F_{\rm H\alpha,0.8~R_e}$ and $F_{\rm H\alpha,1.5~R_e}$ are the cumulative H$\alpha$ flux inside $0.8$ and $1.5$ $r-$band effective radius, respectively. We find that $C_{\rm H\alpha}$ is strongly correlated with the luminosity weighted stellar age gradient. $C_{\rm H\alpha}$ is also sensitive to environmental effects, in the sense that low-mass satellite galaxies below the star formation main sequence tend to have higher $C_{\rm H\alpha}$. For central galaxies, we find that massive disk galaxies with enhanced star formation rate tend to have higher $C_{\rm H\alpha}$, while such a phenomenon is not seen in the low-mass regime. We interpret this as evidence that compaction events more frequently occur in the high-mass regime, which eventually resulting in the buildup of prominent bulges in massive SFGs. Implications of these findings on galaxy structure formation are discussed.
\end{abstract}

\keywords{galaxies: evolution}

\section{Introduction} \label{sec:intro}
The spatially resolved star formation properties of galaxies play a crucial role in understanding how galaxies grow and evolve. In the literature, rest-frame ultraviolet (UV) and H$\alpha$ flux are two most frequently used star formation indicators, tracing star formation rates (SFRs) on timescales of $<100~$Myrs and $<20~$Myrs, respectively \citep{Kennicutt 1998}. Observationally, interstellar dust absorbs starlight and introduce large uncertainties in SFR measurements, especially in the UV band ($\sim 1300-2500$ \AA). To obtain attenuation-corrected SFRs, infrared (IR) data are needed, as starlight removed at short wavelengths by the interstellar dust is re-radiated in the IR \citep{Calzetti 2007, Kennicutt 2009,Hao 2011}. Based on spatially resolved UV-optical-IR photometric data, various topics such as the star formation law \citep{Bigiel 2008, Wyder 2009, Sun 2023}, star formation quenching \citep{Pan 2014,Pan 2016}, spatially resolved star formation main sequence \citep{Abdorrouf 2017,Enia 2020} are explored.

\begin{figure*}
\centering
\includegraphics[width=160mm,angle=0]{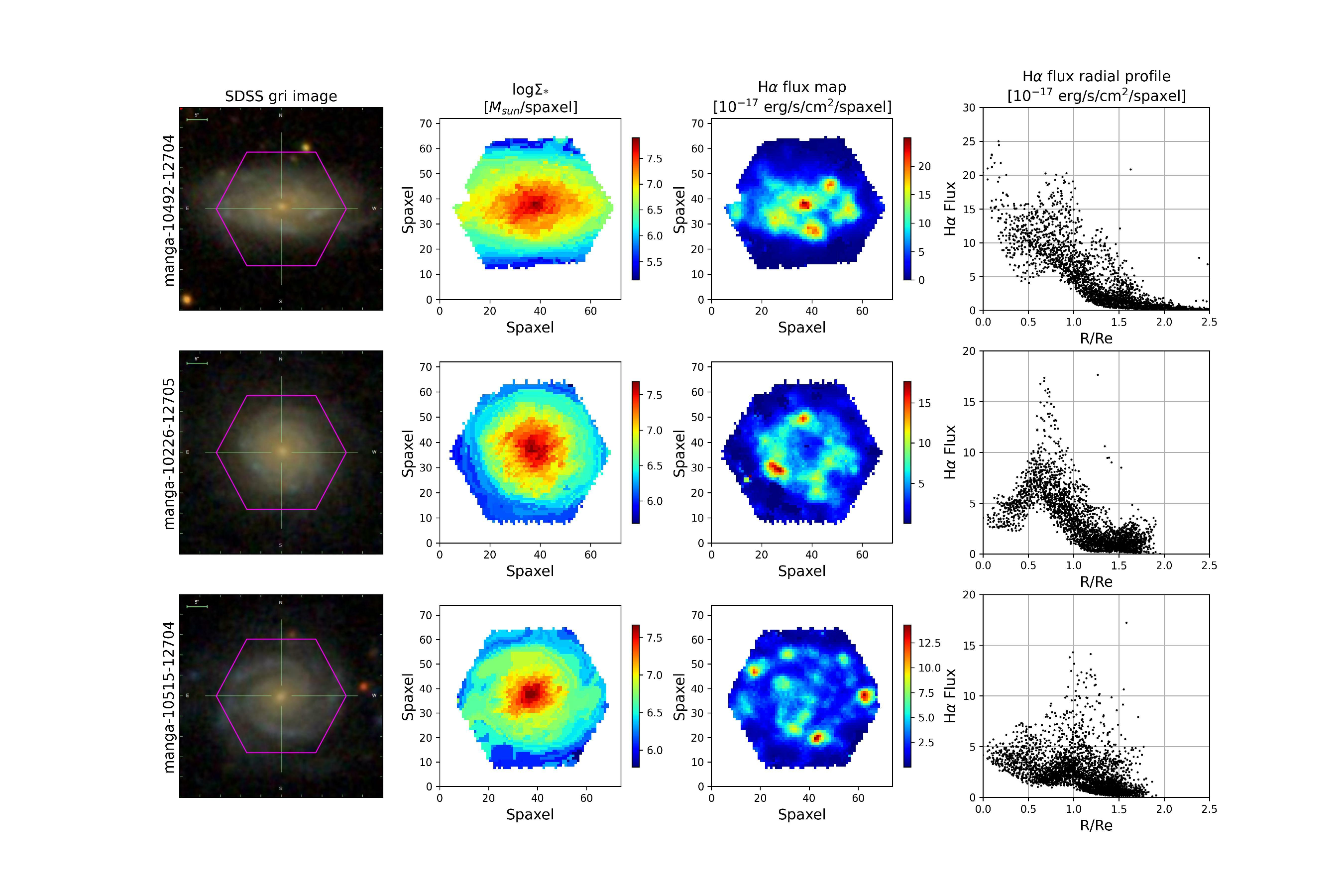}
\caption{Examples of the observed stellar mass and H$\alpha$ maps of MaNGA SFGs. Each row represents a different galaxy, labeled with its MaNGA plate-ifu name. From left to right, the columns represent (1) the SDSS $gri$ composite image, with MaNGA bundle superposed, (2) the stellar mass surface density map, (3) the H$\alpha$ flux (with dust correction applied) surface density map, (4) the H$\alpha$ flux of individual spaxel versus radial distance. As shown in this figure, the spatial distribution of H$\alpha$ flux show clumpy structures, which is far more irregular than that of stellar mass for individual galaxies.}\label{fig1}
\end{figure*}

In the past decade, the advent of large integral field spectroscopy (IFS) surveys [such as CALIFA, \citet{Sanchez 2012}, MaNGA, \citet{Bundy 2015}, SAMI, \citet{Bryant 2015} ] enables more in depth studies on the resolved star formation properties of galaxies. With spectroscopic data, dust correction can be more accurately estimated based on emission line diagnostics such as the Balmer decrement. There have been many studies focusing on the spatially resolved star formation properties at $z<0.2$ \citep[e.g.,][]{Perez 2013,CanoDiaz 2016,Ellison 2018,Bluck 2020}. Among these works, the SFR (or specific star formation rate, sSFR) radial profiles are widely used to characterize the star formation distribution of galaxies  \citep{Gonzalez 2016,Schaefer 2017, Belfiore 2018,Medling 2018, Wang 2018,Wang 2019,Tang 2020}. Some studies presented the SFR radial gradients derived from modeling the SFR radial profiles with linear fit \citep{Schaefer 2017,Wang 2019, Sanchez 2022}. Physically, star formation is triggered through the collapse of giant molecular clumps. Thus star-forming regions often exhibit clumpy structures in galaxies (also see Figure~\ref{fig1}). In this regard, a linear fit to the SFR radial profiles, as done in some previous works, may provide a poor description to the SFR spatial distribution.

The spatial distribution of star formation properties also provide valuable insights to understand the structure formation of galaxies. At $z=2-3$, previous works have revealed that a significant population of disk galaxies host highly dust obscured cores, indicating that these galaxies are rapidly building up dense bulges via central starburst events \citep{Tadaki 2017,Fujimoto 2018}. Recently, the spatially-resolved H$\alpha$ studies on normal SFGs near the star formation main sequence have greatly improved our understanding of galaxy structure formation at $z>2$ \citep{Wilman 2020,Math 2024, Shen 2024}.  Using a sample of $\sim 3600$ SFGs from MaNGA, \citet{Pan 2023} showed that the fraction of SFGs with positive stellar age gradient peak near log$(M_{\ast}/M_{\sun})=10$, indicating that local SFGs build up central mass most actively in this mass regime.

In this work, we aim to investigate the star formation distribution of galaxies through their H$\alpha$ concentrations, and link the H$\alpha$ concentration to the on-going structure formation of SFGs based on the SFG sample compiled by \citet{Pan 2023}. Using the SAMI IFS data, \citet{Schaefer 2017} and \citet{Wang 2022} have also investigated the star formation concentration index of SFGs, with the index defined as $C_{\rm index}={\rm log}(r_{\rm 50,H\alpha}/r_{\rm 50,cont})$, where $r_{\rm 50,H\alpha}$ and $r_{\rm 50,cont}$ are the half light radius of H$\alpha$ flux and $r-$band continuum, respectively. In this work, we will define the star formation concentration index with a simpler approach. Throughout this paper, we adopt a concordance $\Lambda$CDM cosmology with $\Omega_{\rm m}=0.3$, $\Omega_{\rm \Lambda}=0.7$, $H_{\rm 0}=70$ $\rm km~s^{-1}$ Mpc$^{-1}$ and a \citet{Chabrier 2003} initial mass function (IMF).

\section{the sample}
MaNGA is the largest existing IFS survey for nearby galaxies, which has gathered spatially resolved spectroscopic information for $\sim 10,000$ galaxies at $z<0.15$ \citep{Bundy 2015}. MaNGA uses a series of hexagonal optical fiber bundles with different sizes (varying from 19 to 127 fibers) to cover the target galaxies, which corresponds to a field of view (FoV) of $12\arcsec .5-32\arcsec .5$ in diameter \citep{Law 2015}. The MaNGA sample is composed of the "Primary" and the "Secondary" sample, for which the radial coverage reaches at least out to $\sim 1.5~R_{\rm e}$ and $\sim 2.5 ~R_{\rm e}$ \footnote{$R_{\rm e}$ is the half light radius measured from the SDSS $r-$band imaging data, which is drawn from the NASA-sloan Atlas (htts://nsatlas.org).}, respectively. In this work, both samples are used. The MaNGA data productions have been released in SDSS data release 17 (DR17) \footnote{https://www.sdss4.org/dr17/manga/}.

Our SFG sample is compiled by \citet{Pan 2023}, which is exacted from the MaNGA Pipe3D value-added catalog \citep{Sanchez 2022}.  The detailed sample selection criteria are as follows.

1. Stellar mass $10^{9.0}M_{\sun}<M_{\ast}<10^{12}M_{\sun}$;

2. Redshift range $0.01<z<0.15$;

3. Minor-to-major axis ratio $b/a>0.4$ to exclude edge on objects;

4. Specific star formation rate log($\rm sSFR/yr^{-1})>-11.0$, where sSFR=SFR/$M_{\ast}$;

5. Integral filed unit (IFU) size (i.e., the fiber number of the fiber bundle) $>$ 19 to exclude galaxies with possibly insufficient resolution;

6. Quality control flag QCFLAG = 0 to exclude data cubes with bad qualities.

These selection criteria yield 3590 SFGs. To ensure that the H$\alpha$ concentration can be robustly measured, we visually inspect the H$\alpha$ map of each source via the map file produced by the MaNGA data analysis pipeline (DAP) \citep{Belfiore 2019,Westfall 2019}. When a source has a significant number of masked bad spaxels in its H$\alpha$ map, we reject the source from further analysis. This procedure removes 492 galaxies. Our final sample contains 3098 SFGs.

One may consider whether the selected MaNGA SFG sample is representative for the local SFG population. In \citet{Pan 2023}, we have showed that at log$(M_{\ast}/M_{\odot})>10.0$, our selected MaNGA SFG sample has a similar distribution as the SFG sample selected at $z=[0.02,0.05]$ in the $R_{\rm e}-M_{\ast}$ plane. At log$(M_{\ast}/M_{\odot})<10.0$, however, the MaNGA SFG sample is biased against compact galaxies (see figure 5 of \citealt{Pan 2023}). This is mainly because low-mass compact galaxies are difficult to spatially resolved due to their small physical sizes, thus such sources are less selected for the MaNGA observation. In addition, the exclusion of galaxies with small IFU sizes (the fourth bullet point) somewhat contributes to this bias. However, the main conclusion of this work should be insensitive to this sample selection effect since we focus on disk galaxies, which have relatively extended morphologies.

\begin{figure*}
\centering
\includegraphics[width=140mm,angle=0]{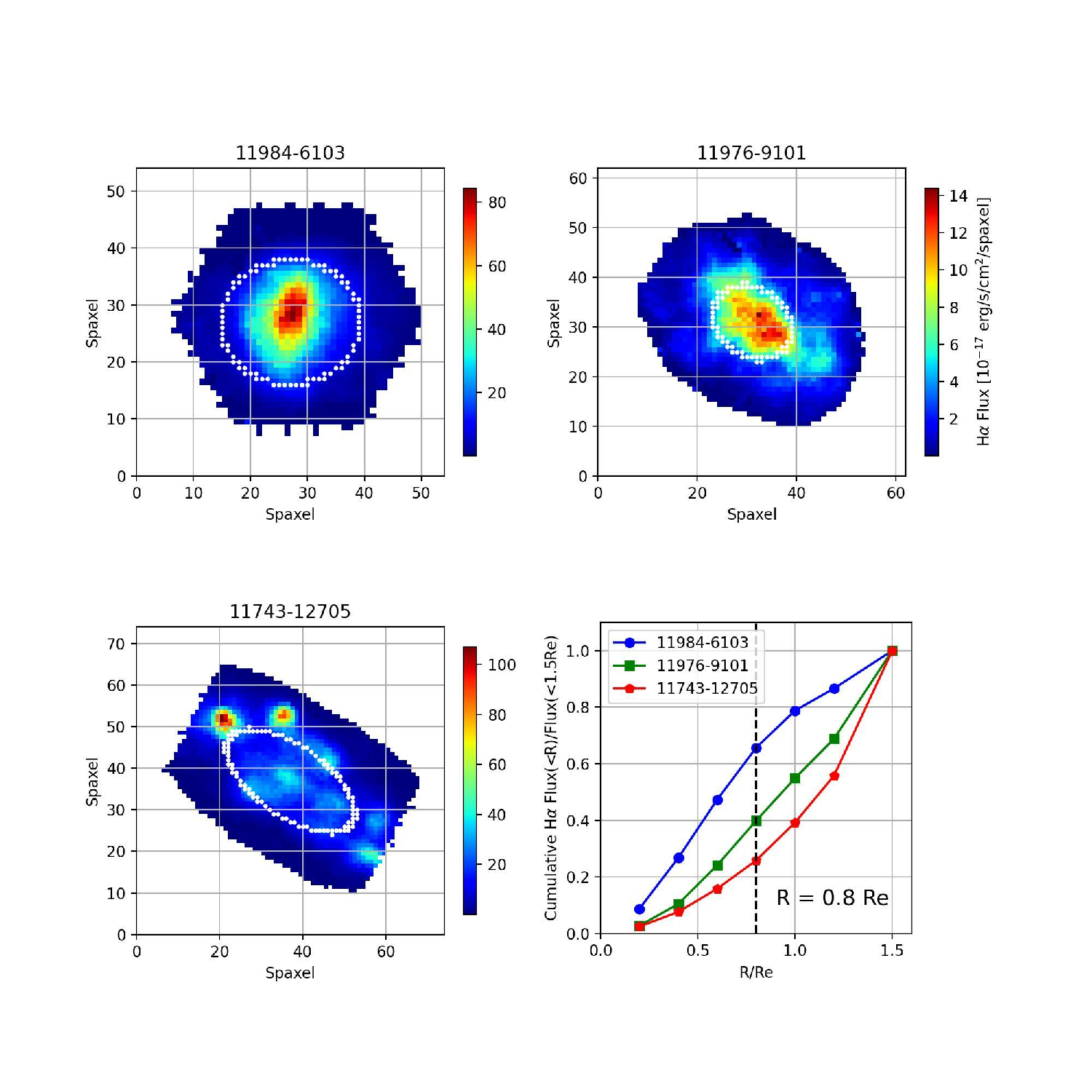}
\caption{Examples of 3 H$\alpha$ maps and their H$\alpha$ flux growth curves. In each H$\alpha$ map, the white dots indicate spaxels with $R=[0.95,1.05]~R_{\rm e}$. The H$\alpha$ flux growth curves of the 3 galaxies are shown in the bottom right panel. The dashed line marks the radius of $R=0.8~R_{\rm e}$.}\label{fig2}
\end{figure*}

\section{Measuring the H$\alpha$ concentration}
In this work, we aim to measure the H$\alpha$ flux concentration index ($C_{\rm H\alpha}$) for the MaNGA SFG sample. In most cases, the FoV does not cover the whole body of the target galaxy. We thus do not define the concentration as that used in imaging surveys. In the CAS system (concentration, asymmetry, clumpiness; \citealt{Conselice 2003}), the concentration index is defined as $C=R_{80}/R_{20}$, where $R_{80}$ and $R_{20}$ is the radius that enclosing 80\% and 20\% of the total flux, respectively. In the SDSS studies, the concentration index is defined as $R_{90}/R_{50}$, where $R_{90}$ and $R_{50}$ is the radius that enclosing 90\% and 50\% of the petrosian flux, respectively \citep{Strateva 2001,Kauffmann 2003}. In the MaNGA survey, the "Primary" and "Secondary" sample have different spatial resolutions. To enable a uniform $C_{\rm H\alpha}$ measurement for both samples, we define the H$\alpha$ concentration index based on the H$\alpha$ flux distribution within $1.5~R_{\rm e}$.

We use the growth curve method to define $C_{\rm H\alpha}$. Before measuring the H$\alpha$ growth curve, dust attenuation correction must be applied to the observed H$\alpha$ flux. We use the Balmer decrement method to quantify the dust attenuation in star-forming (SF) spaxels. For SF regions, the H$\alpha$ emission is produced by the recombination of gas ionized by young stars. First, we employ the BPT diagram \citep{Baldwin 1981, Kauffmann 2003} to classify the MaNGA spaxels in to SF, composite and syfert classes. For the SF class, we use the equation of \citet{DS 2013} to derive the color excess $E(B-V)$:
\begin{equation}
E(B-V)=1.97{\rm log}_{10}[\frac{({\rm H}\alpha/{\rm H}\beta)_{\rm obs}}{2.87}]
\end{equation}
In this equation, an intrinsic value for $({\rm H}\alpha/{\rm H}\beta)_{\rm int}$ is assumed to be 2.87, under Case B recombination conditions, with an electron temperature of $T_{\rm e}=10^{4}$ K and an electron density of $n_{\rm e}=10^{2} cm^{-3}$.

For SF spaxels, the intrinsic H$\alpha$ flux
$f_{\rm {int, H\alpha}}$ is determined by:
\begin{equation}
f_{\rm {int,H\alpha}}=f_{\rm {obs,H\alpha}}\times10^{0.4A_{\rm H\alpha}},
\end{equation}
where the observed H$\alpha$ flux $f_{\rm {obs,H\alpha}}$ is from the DAP, and $A_{\rm H\alpha}=2.38E(B-V)$ is the H$\alpha$ dust attenuation (using the \citet{Calzetti 2000} reddening curve with $R_{v}=3.1$).

\begin{figure*}
\centering
\includegraphics[width=160mm,angle=0]{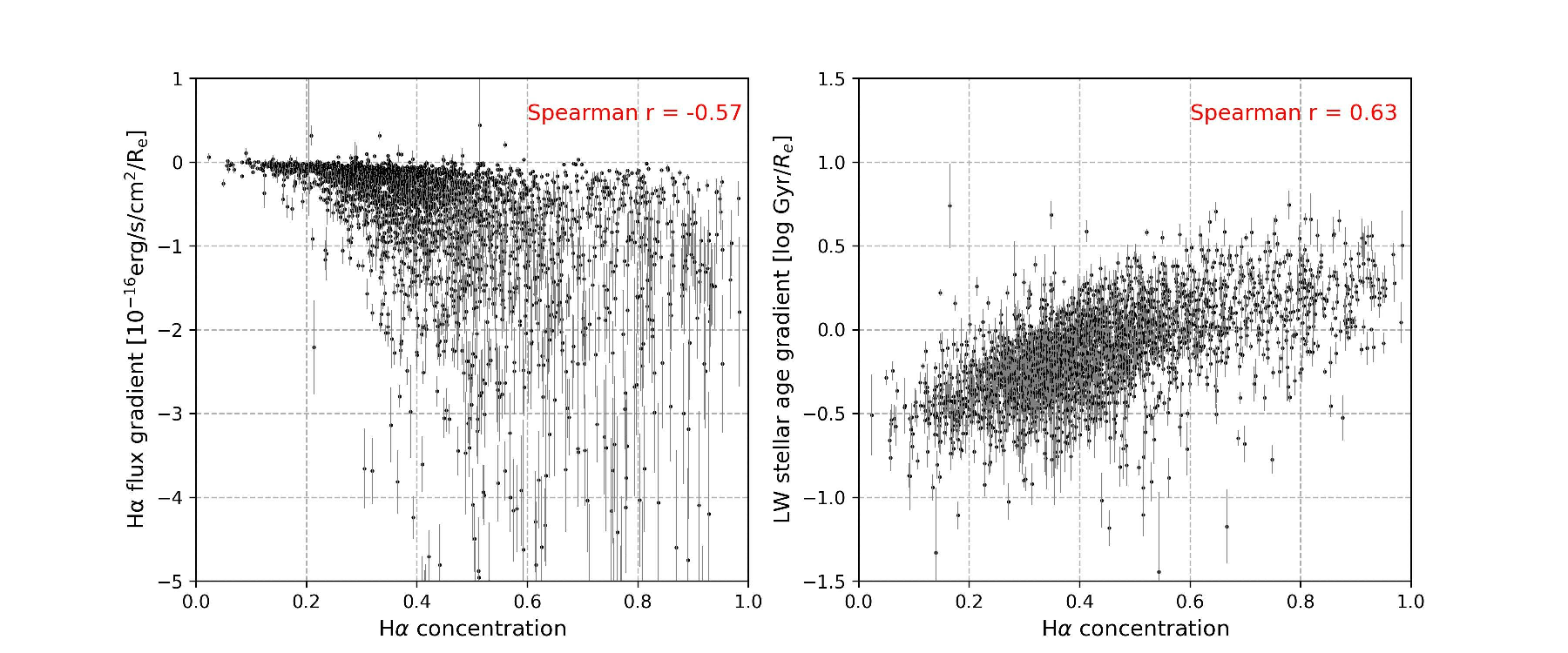}
\caption{Left: The H$\alpha$ radial gradient as a function of H$\alpha$ concentration. The Spearmann coefficient is marked in the top right corner. Right: The light weighted stellar age gradient as a function of H$\alpha$ concentration. }\label{fig3}
\end{figure*}

\begin{figure*}
\centering
\includegraphics[width=160mm,angle=0]{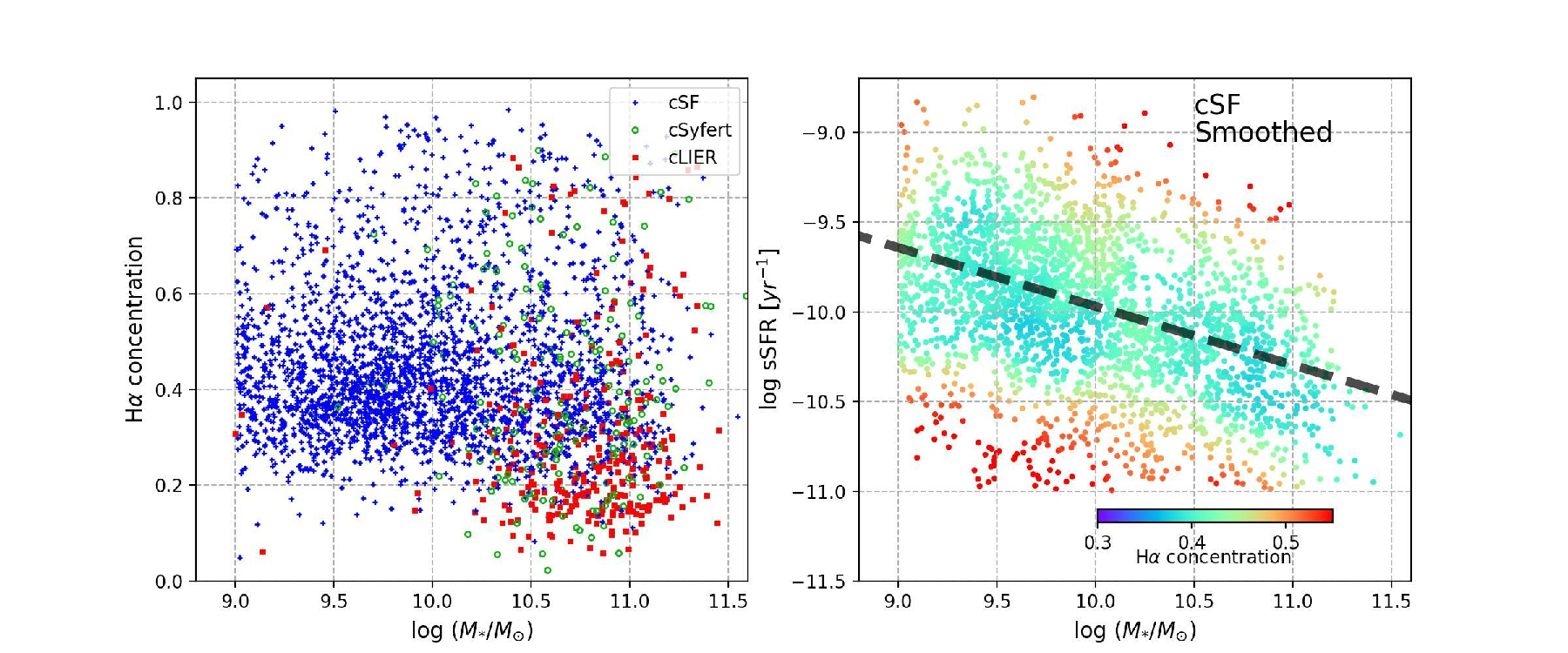}
\caption{Left: H$\alpha$ concentration as a function of stellar mass. Different symbols indicate 3 categories of galaxies classified by the $\rm S_{\rm II}-$BPT diagram, based on the line ratios of the central $2\arcsec .5$-diameter aperture. Right: the sSFR$-M_{\ast}$ diagram of the galaxies classified as the cSF type. Symbols are color coded by the H$\alpha$ concentration. The dashed line indicate a linear fitting of the symbols, i.e, the SFMS. To show the trend clearly, a LOESS smoothing method is applied.}\label{fig4}
\end{figure*}

We notice that we do not apply dust attenuation correction to spaxels classified as composite or syfert classes. For massive SFGs ($M_{\ast}>10^{10}M_{\sun}$), the central regions of many galaxies fall in these two classes (also see Figure~\ref{fig4}). Given this, we classified our sample into 3 categories based on the line ratios of the central $2\arcsec .5$-diameter aperture as measured by the Pipe3D pipleline using the $\rm S_{\rm II}-$BPT diagram. The conclusions of this work are based on the cSF sample, for which the central $2\arcsec .5$-aperture regions are also classified as the SF class. Thus our conclusions should be robust against dust correction applied.

The H$\alpha$ emission is not only from star-forming regions (i.e., the \hii~regions) and active galactic nuclei regions, but can also be from the diffuse ionized gas (DIG)\citep{Madsen 2006,Zhang 2017,Mc 2024}. The H$\alpha$ emission of DIG can be powered by the leaking radiation from \hii~regions, or by the hot evolved stars. In a recent work, \citet{Belfiore 2022} suggest that in $z\sim0$ SFGs, the leaking radiation from \hii~regions is the dominant contributor to the DIG energy budget, while the contribution from hot evolved stars is only $\sim2$\%. The most part of the following analysis is focused on the cSF sample, it is thus safe to treat $C_{\rm H\alpha}$ as an indicator of star formation spatial distribution.

We then measure the growth curve of dust-corrected H$\alpha$ flux out to $1.5~R_{\rm e}$, with a bin width of $0.2~R_{\rm e}$. The $R/R_{\rm e}$ for each spaxel is taken from the DAP map file, which has taken ellipticity into account. Some of the examples are shown in Figure~\ref{fig2}. As can be seen, galaxies with visually different H$\alpha$ concentrations have different H$\alpha$ growth curves. In this work, we define the H$\alpha$ concentration index $C_{\rm H\alpha}$ as:
\begin{equation}
C_{\rm H\alpha}=F_{\rm H\alpha,0.8~R_e}/F_{\rm H\alpha,1.5~R_e},
\end{equation}
where $F_{\rm H\alpha,0.8~R_e}$ and $F_{\rm H\alpha,1.5~R_e}$ is the cumulative H$\alpha$ flux within 0.8 $R_{\rm e}$ and 1.5 $R_{\rm e}$, respectively. We choose an aperture of $R=0.8~R_{\rm e}$ to define H$\alpha$ concentration because: (1) this aperture is large enough to avoid the point spread function (PSF) smoothing effects on the $C_{\rm H\alpha}$ measurement. By design, the spatial FWHM (i.e., the resolution) is $\sim 0.4~R_{\rm e}$ and $\sim 0.6~R_{\rm e}$ for the "Primary" and "Secondary" sample, respectively \citep{Bundy 2015}; (2) at $R=0.8~R_{\rm e}$, different growth curves can be well separated. As shown in Figure~\ref{fig2}, growth curves can not be well separated when the radius is too small or too large. In the Appendix, we also check the dependence of $C_{\rm H\alpha}$ on resolution. We show that when $M_{\ast}$ and Sersic index $n$ are fixed, the measured $C_{\rm H\alpha}$ shows no dependence on resolution.

For spaxels with $R<1.5~R_{\rm e}$, the typical signal-to-noise ratio (S/N) of H$\alpha$ flux is around $5-50$ \citep{Belfiore 2019}. By definition, $F_{\rm H\alpha,0.8~R_e}$ and $F_{\rm H\alpha,1.5~R_e}$ are the integrated H$\alpha$ fluxes of many spaxels ($N_{\rm spaxel}\sim 10^{2}-10^{3}$). Thus they have very high S/Ns (~$\times \sqrt{N_{\rm spaxel}}$ higher than that of individual spaxel).  The uncertainty of $C_{\rm H\alpha}$ is typically $<1$\%. In the following plots, we do not overplot the errorbar of $C_{\rm H\alpha}$ on the symbol since it is typically smaller than the symbol marked.

\begin{figure*}
\centering
\includegraphics[width=140mm,angle=0]{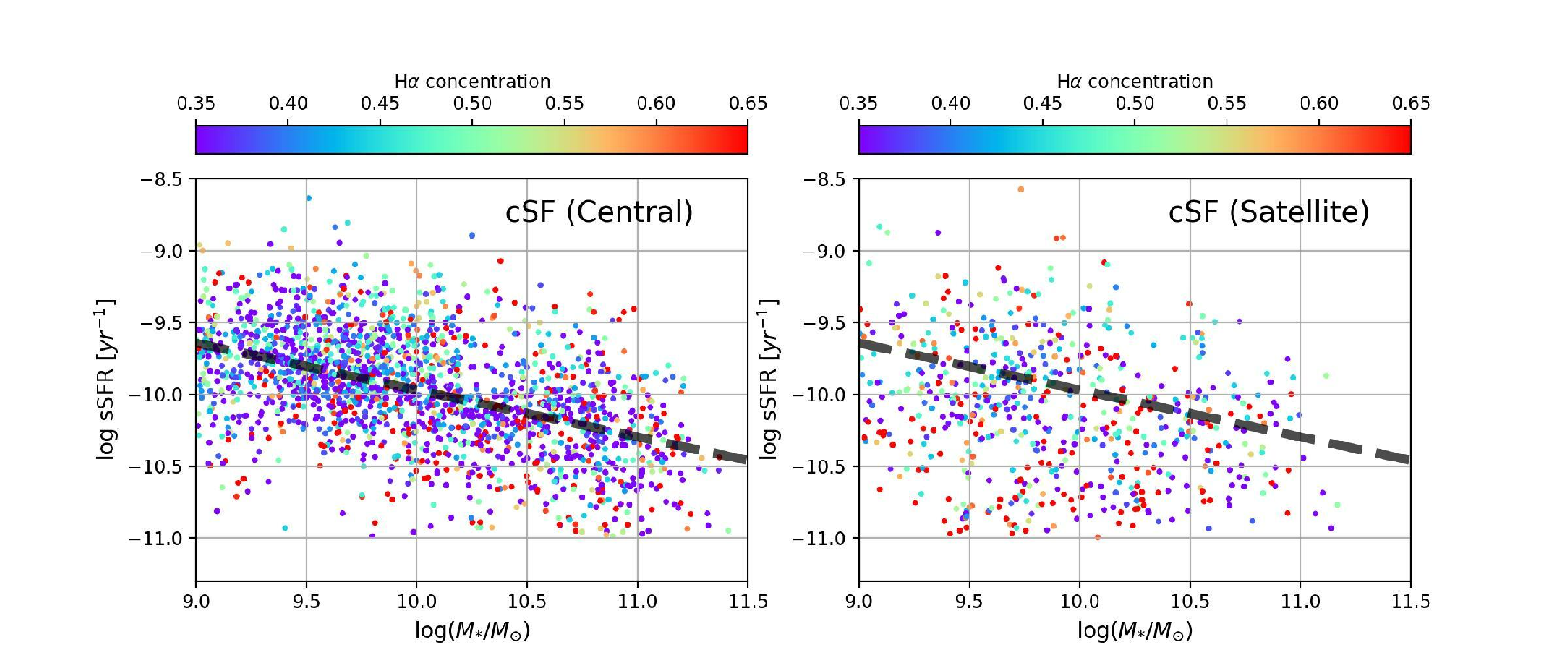}
\caption{Left: the sSFR$-M_{\ast}$ relation for central galaxies that are classified as the cSF type. Right: Similar to the left panel, but for the satellite galaxies. In each panel, the dashed line shows the SFMS.}\label{fig5}
\end{figure*}

\section{Results}
\subsection{Comparisons among H$\alpha$ concentration, H$\alpha$ flux gradient and stellar age gradient}
The H$\alpha$ flux radial gradient is widely used to characterize the spatial distribution of H$\alpha$ in IFS studies. In this section we first compare the H$\alpha$ concentration with the H$\alpha$ flux gradient drawn from the MaNGA Pipe3D catalog. In Pipe3D, the H$\alpha$ flux gradient is computed as the slope of the linear fit to the H$\alpha$ flux radial profile. The gradient is derived based on a linear regression, and the fitting is restricted at $0.5-2.0~R_{\rm e}$. When the FoV does not reach 2.0~$R_{\rm e}$, the regression is restricted to the largest distance covered by the FoV.

Figure~\ref{fig3} shows the comparison between H$\alpha$ concentration and H$\alpha$ gradient. Overall, galaxies with low $C_{\rm H\alpha}$ have shallow H$\alpha$ flux gradients, but the correlation between these two quantities exhibits a large dispersion, especially in the high $C_{\rm H\alpha}$ regime. A Spearman coefficient of $r=-0.57$ is found between these two quantities. We attribute the relatively poor correlation between these two quantities to two aspects. (1) The H$\alpha$ flux gradient is poorly constrained for many sources. As can be seen in Figure~\ref{fig3}, the H$\alpha$ flux gradient has a large uncertainty for a significant fraction of sources. This is natural since the distribution of H$\alpha$ is quite clumpy, as illustrated in Figure~\ref{fig1}. (2) The radius range adopted in the H$\alpha$ gradient fitting is $0.5-2.0~R_{\rm e}$. As shown in previous studies, the radial profile of a physical property in the central region of a galaxy is smoothed by the PSF and becomes flattened compared to the intrinsic one \citep{Belfiore 2017b}. Thus the Pipe3D pipeline does not include regions with $R<0.5~R_{\rm e}$ in the gradient fitting. In fact, the $R<0.5~R_{\rm e}$ region contribute $\sim10-30$\% to $F_{\rm H\alpha,1.5~R_e,}$ (see Figure~\ref{fig2}).

In the right panel, we compare the H$\alpha$ concentration with the luminosity-weighted (LW) stellar age gradient. The Pipe3D pipeline uses a library which compromises 273 simple stellar population (SSP) spectra, sampling 39 ages from 1 Myr to 13.5 Gyr and seven metallicities ($Z/Z_{\sun}$=0.006, 0.029, 0.118, 0.471, 1, 1.764 and 2.353), to model the observed spectra and derive the luminosity-weighted and stellar mass-weighted stellar age \citep{Sanchez 2022}. Surprisingly, $C_{\rm H\alpha}$ is quite strongly correlated with the LW stellar age gradient, with a Spearman coefficient of $r=0.63$. This suggests that the current star formation distribution of a galaxy is coupled with its formation history. In summary, Figure~\ref{fig3} encourages us to believe that H$\alpha$ concentration is a proper parameter in characterizing the 2D star formation distribution of galaxies.

\subsection{The distribution of H$\alpha$ concentration across the star-forming population}
The left panel of Figure~\ref{fig4} presents H$\alpha$ concentration as a function of stellar mass $M_{\ast}$. Based on the line ratios of the central $2\arcsec .5$-diameter aperture provided by the Pipe3D catalog, we first classify our SFG sample into "cSF", "cLIER" and "cSyfert" types using the $\rm S_{\rm II}-$BPT diagram. SFGs of different types are shown in different symbols in the figure. As shown in Figure~\ref{fig4}, the majority of cSF galaxies have $C_{\rm H\alpha}\sim 0.4$. At log$(M_{\ast}/M_{\sun})>10.3$, a significant fraction of SFGs are classified as cLIERs. A cLIER galaxy is typically composed by an old bulge+a star-forming disk component, which is considered to be quenching from the inside-out \citep{Belfiroe 2016, Belfiore 2017}. As can be seen, at fixed $M_{\ast}$, cLIERs generally have lower $C_{\rm H\alpha}$ than the cSF ones, consistent with the "inside-out" star formation quenching scenario. Some cLIERs also exhibit high $C_{\rm H\alpha}$. Such sources are likely low-luminosity AGNs. Syfert galaxies distribute similarly as cLIERs in this diagram. Since we do not apply dust correction to the non-SF spaxels, in the following sections, we focus our analysis on the cSF class.

\begin{figure*}
\centering
\includegraphics[width=140mm,angle=0]{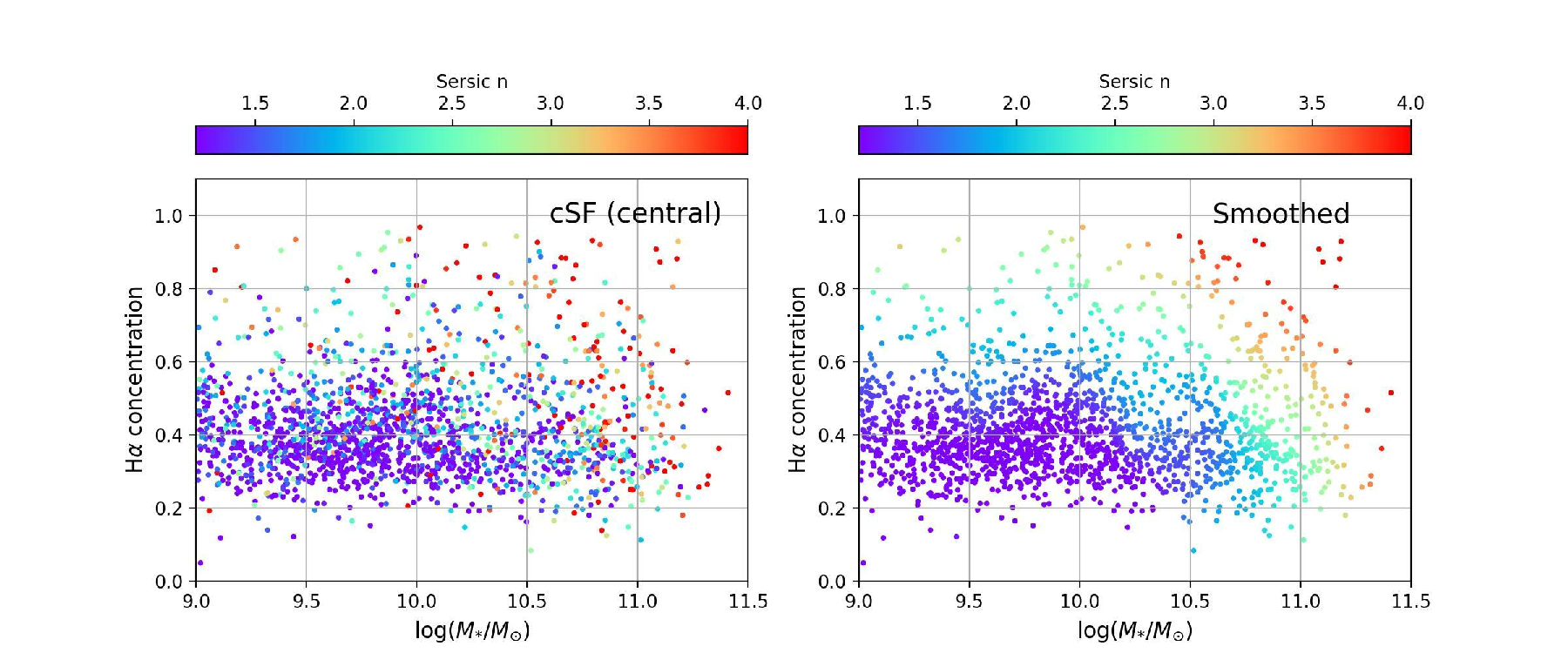}
\caption{Left: The H$\alpha$ concentration as a function of stellar mass for the central galaxies with the cSF type. Symbols are color coded by Sersic index $n$. Right: a LOESS smooth version of the left panel. }\label{fig6}
\end{figure*}

\begin{figure*}
\centering
\includegraphics[width=140mm,angle=0]{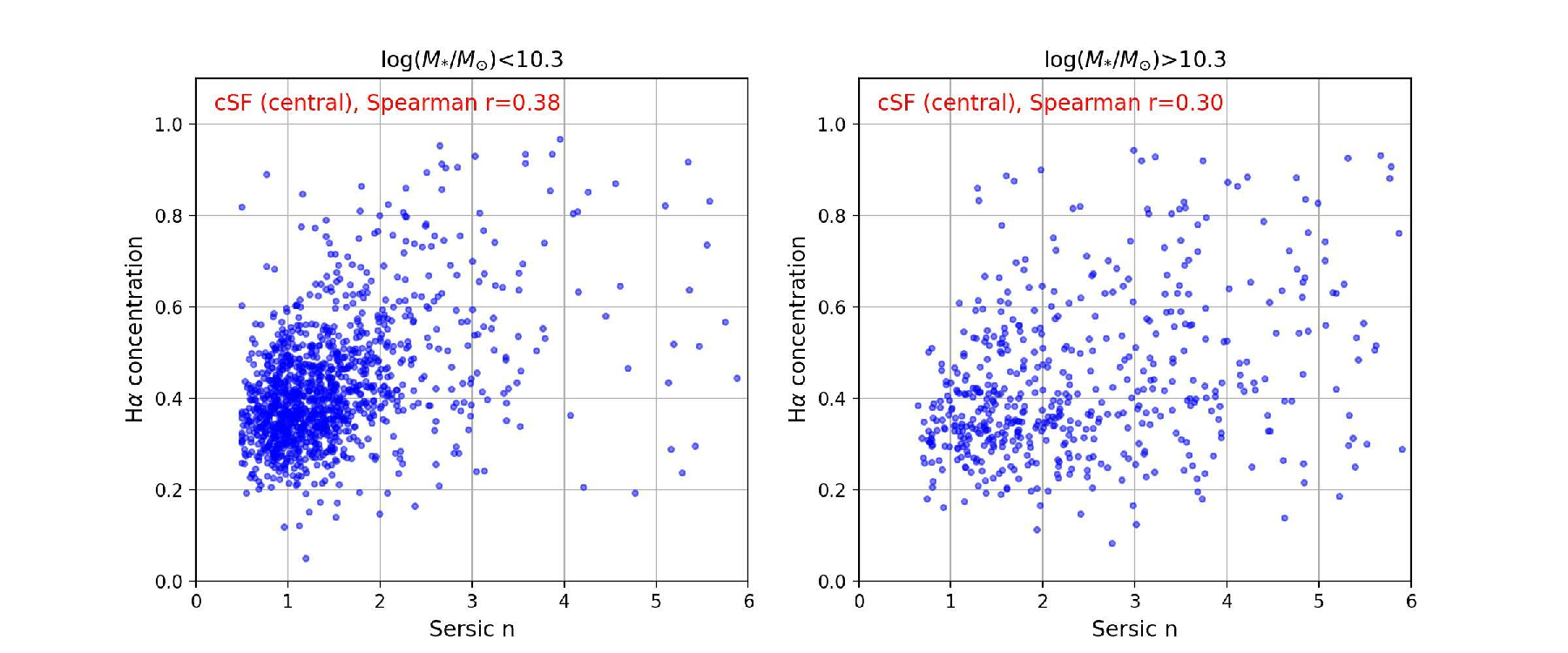}
\caption{Left: The H$\alpha$ concentration as a function of Sersic index $n$ for central galaxies with the cSF type for the log$(M_{\ast}/M_{\sun})<10.3$ mass bin. Right: similar to the left panel, but for the log$(M_{\ast}/M_{\sun})>10.3$ mass bin  }\label{fig7}
\end{figure*}

In the right panel of Figure~\ref{fig4}, we show the sSFR$-M_{\ast}$ relation for the cSF class. The star formation main sequence (SFMS) of local SFGs is shown in the dashed line \citep{Renzini 2015}. Symbols are color-coded by H$\alpha$ concentration. To show the H$\alpha$ concentration distribution more clearly, the data are smoothed using the LOESS method developed by \cite{Cleveland 1988}, with the \verb"PYTHON" code provided by \cite{Cappellari 2013}. As can be seen, galaxies near the SFMS ridge line exhibit the lowest H$\alpha$ concentration, while those in the upper and lower envelopes of the SFMS show enhanced $C_{\rm H\alpha}$. Such a trend is quite similar to the galaxy Sersic index distribution across the SFMS, as reported in early studies \citep{Wuyts 2011}. In the following sections, we will come back to this issue and discuss its physical origin.

\subsection{The dependence of H$\alpha$ concentration on environment}
The environmental effects have significant influences in shaping the star formation properties of galaxies. In this section, we explore the dependence of H$\alpha$ concentration upon environment.

We classify the cSF class galaxies into central and satellite galaxies based on the SDSS group catalog of \citet{Yang 2007}. As shown in previous studies, the star formation properties of low-mass satellite galaxies are sensitive to environmental effects \citep{Peng 2012}. In observations, environmental effects can cause "outside-in" star formation quenching in low-mass galaxies. Based on the SAMI sample, \citet{Schaefer 2017} showed that galaxies in high density environment have steepened H$\alpha$ slopes, i.e., their star formation distributions are more centrally concentrated. In Figure~\ref{fig5}, we show the sSFR$-M_{\ast}$ relation separately for central and satellite galaxies. Symbols are coded by $C_{\rm H\alpha}$. A most remarkable feature is that, in the lower envelope of the SFMS, low-mass satellite galaxies have higher $C_{\rm H\alpha}$. This is nicely in line with the picture of environment-driven "outside-in" star formation quenching for low-mass satellite galaxies.

The agreement between our results and previous findings again suggests that H$\alpha$ concentration is a robust indicator for characterizing the star formation distribution of galaxies. To avoid environmental effects in our analysis, in the following sections we only focus on cSF central galaxies.

\begin{figure*}
\centering
\includegraphics[width=140mm,angle=0]{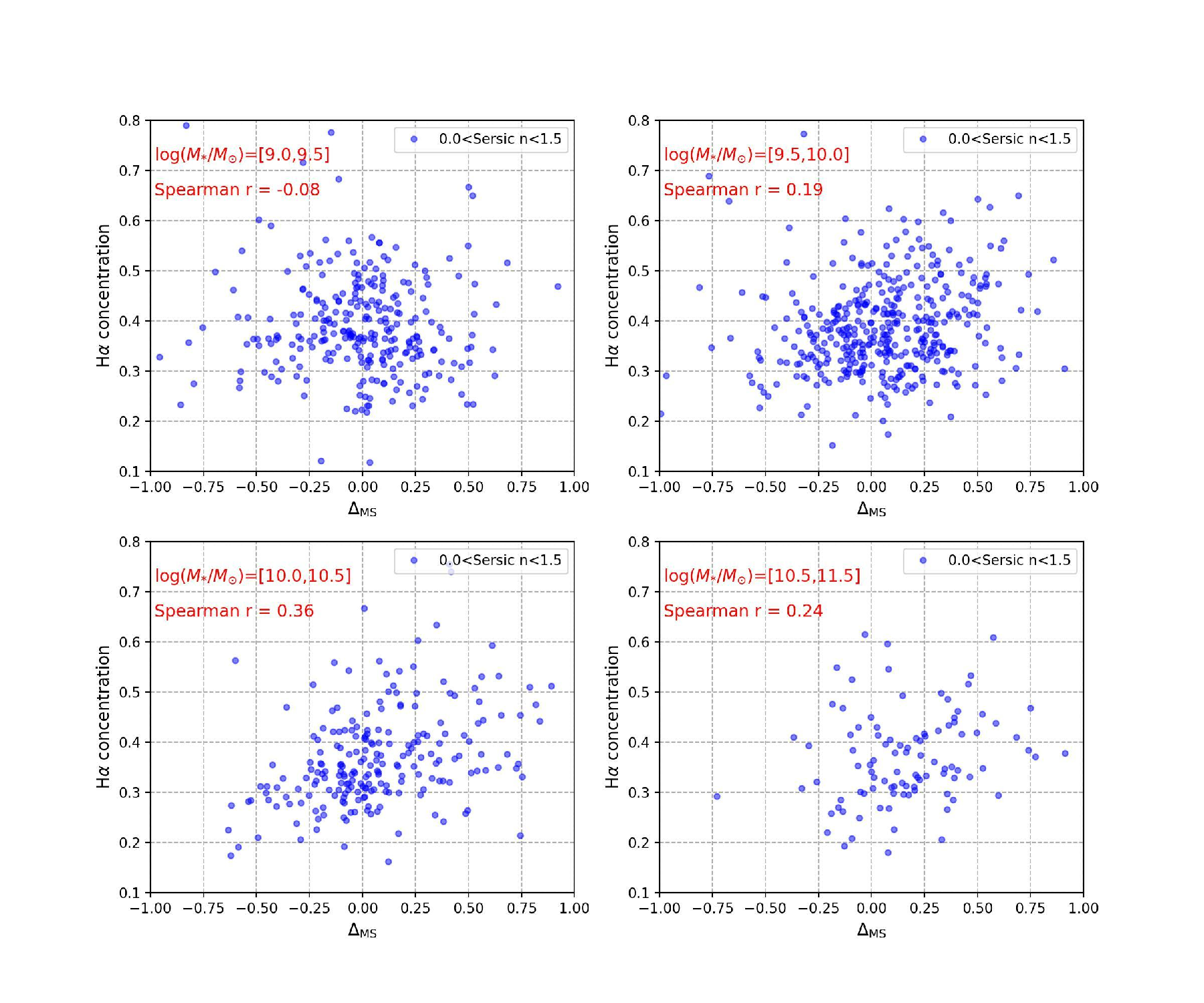}
\caption{H$\alpha$ concentration as a function of $\Delta_{\rm MS}$ for central galaxies with cSF type and $n<1.5$. The samples are divided into 4 stellar mass bins, as shown in different panels. The Spearman correlation coefficient is marked in each panel.  }\label{fig8}
\end{figure*}

\subsection{The dependence of H$\alpha$ concentration on galaxy morphology}
In Figure~\ref{fig6}, we show the $C_{\rm H\alpha}-M_{\ast}$ relation for cSF central galaxies. Symbols are color coded by Sersic index $n$. As can be seen, galaxies with $C_{\rm H\alpha}>0.6$ tend to have high $n$. In the $0.2<C_{\rm H\alpha}<0.6$ regime where most galaxies reside, galaxies with log$(M_{\ast}/M_{\sun})<10.3$ have $n<1.5$, i.e, they are pure disk galaxies. At log$(M_{\ast}/M_{\sun})>10.3$, more massive galaxies tend to have higher Sersic indices. Inspecting the LOESS smoothed version shown in the right panel of Figure~\ref{fig6}, it seems that the morphologies of disk galaxies transform from disk-dominated systems into bulge+disk systems near log$(M_{\ast}/M_{\sun})=10.3$. Nevertheless, we note that there are still some pure disk galaxies in the log$(M_{\ast}/M_{\sun})>10.5$ mass regime.

Figure~\ref{fig7} shows the $C_{\rm H\alpha}-n$ relations for galaxies with log$(M_{\ast}/M_{\sun})<10.3$ and log$(M_{\ast}/M_{\sun})>10.3$. For the low-mass galaxies, there is a clear positive correlation between $n$ and $C_{\rm H\alpha}$, with a Spearman correlation coefficient of $r=0.38$. Such a positive correlation also exists for high-mass galaxies, but the correlation strength is weakened, with $r=0.30$. Note that a significant fraction of massive galaxies with $n>2$ have $C_{\rm H\alpha}<0.4$. We have checked the BPT classification maps of these galaxies. We found that although they are classified as SF in the central 2.5 arcsecond aperture, the innermost spaxels of these galaxies have already exhibited LIER features. Given this, these galaxies are likely at the early stage transforming from cSFs into cLIERs.

\begin{figure*}
\centering
\includegraphics[width=140mm,angle=0]{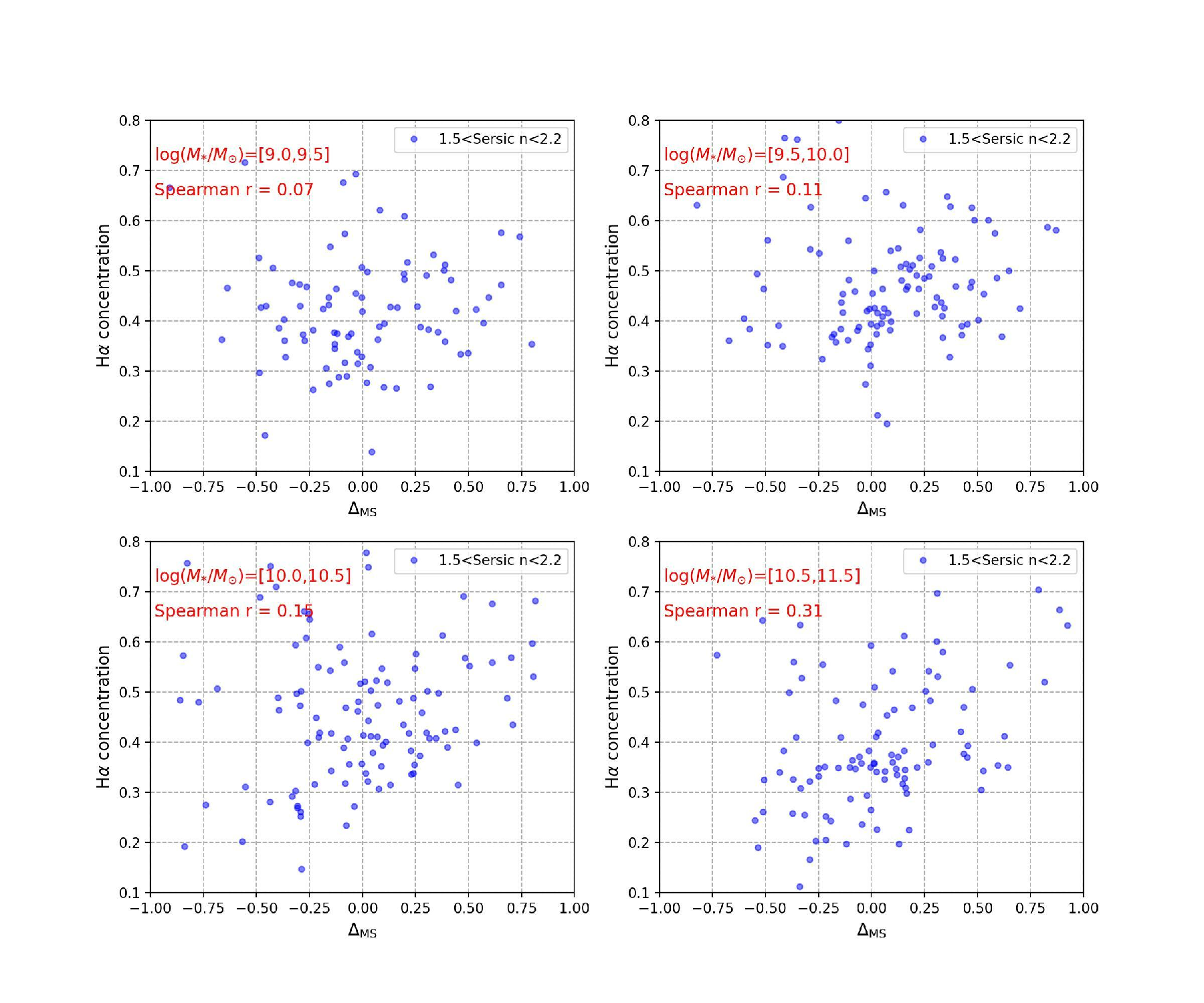}
\caption{Similar to Figure 8, but for galaxies with $1.5<n<2.2$.}\label{fig9}
\end{figure*}

\subsection{The dependence of H$\alpha$ concentration upon SFR for disk galaxies}
The correlation between galaxy structure and $C_{\rm H\alpha}$ explains why SFGs in the upper and lower envelopes of the SFMS have enhanced $C_{\rm H\alpha}$ (see Figure~\ref{fig4}), since in these regions galaxies typically have bulge-dominated morphologies \citep{Wuyts 2011}. In this section, we focus on disk-dominated SFGs. As shown in early works, disk-dominated galaxies likely form their stellar mass in a more elongated time scale and have younger stellar population age \citep{Gonzalez 2015,Mendez 2021}. Investigating the $C_{\rm H\alpha}$ distribution of these galaxies thus can provide insights in understanding their recent structure formation histories.

We first focus on pure disk galaxies (with $n<1.5$). From Figure~\ref{fig6} and Figure~\ref{fig7}, it is clear that galaxies with $n<1.5$ primarily have $0.2<C_{\rm H\alpha}<0.6$, with no significant dependence on $M_{\ast}$. In Figure~\ref{fig8}, we investigate the $C_{\rm H\alpha}$ dependence upon $\Delta_{\rm MS}$, where $\Delta_{\rm MS}$ is offset from the SFMS ridge line (defined as $\Delta_{\rm MS}=\rm log(SFR/SFR_{\rm MS})$). We calculate the Spearman correlation coefficient between the two quantities in 4 stellar mass bins. As shown in Figure~\ref{fig8}, $C_{\rm H\alpha}$ shows no clear dependence on $\Delta_{\rm MS}$ at log$(M_{\ast}/M_{\sun})<9.5$. However, this trend does not hold for massive galaxies. At $10.0<$log$(M_{\ast}/M_{\sun})<10.5$ and log$(M_{\ast}/M_{\sun})>10.5$, the Spearman coefficients are $r=0.36$ and $r=0.24$, indicating that galaxies with higher SFRs tend to have higher $C_{\rm H\alpha}$.

Figure~\ref{fig9} shows the results for disk galaxies with $1.5<n<2.2$. Overall, the trends are similar to those shown in Figure~\ref{fig8}. A slight difference is that the strongest correlation is found in the log$(M_{\ast}/M_{\sun})>10.5$ bin, with $r=0.31$. This may be partially resulted from the different $\Delta_{\rm MS}$ range for galaxies with different $n$ in the log$(M_{\ast}/M_{\sun})>10.5$ bin. As can be seen, in the log$(M_{\ast}/M_{\sun})>10.5$ bin, galaxies with $1.5<n<2.2$ span a wider $\Delta_{\rm MS}$ range, thus resulting in a more pronounced relation between $C_{\rm H\alpha}$ and $\Delta_{\rm MS}$.

%The correlation between $C_{\rm H\alpha}$ and $\Delta_{\rm MS}$ is weakest at log$(M_{\ast}/M_{\sun})<9.5$, with $r=0.07$. A slightly difference is that
%The strongest correlation is found in the log$(M_{\ast}/M_{\sun})>10.5$ bin, with $r=0.31$. In this mass bin, the correlation strength appears stronger than that shown in Figure~\ref{fig8}. This may be due to the relatively narrow $\Delta_{\rm MS}$ range of galaxies with $n<1.5$ bin in the log$(M_{\ast}/M_{\sun})>10.5$ bin. As shown in Figure~\ref{fig8}, massive pure disk galaxies mainly distribute at $\Delta_{\rm MS}>-0.25$. While in the $1.5<n<2.2$ bin, massive galaxies span a wider $\Delta_{\rm MS}$ range, thus resulting in a more pronounced relation between $C_{\rm H\alpha}$ and $\Delta_{\rm MS}$.

Combined together, Figure~\ref{fig8} and Figure~\ref{fig9} suggest that the enhancement of star formation in massive disk galaxies is accompanied by a high star formation concentration. We will discuss this finding in the context of galaxy evolution in the next section.

\section{Discussion}
In this work, we define a parameter, the concentration of H$\alpha$, to characterize the star formation spatial distribution of low$-z$ SFGs. This parameter is model free and easy to measure. We investigate the $C_{\rm H\alpha}$ distribution among low$-z$ SFGs and study its correlations to galaxy structure and SFRs. Below we discuss the implications of our findings.

We show that $C_{\rm H\alpha}$ is quite well correlated with the luminosity weighted stellar age gradient. As stellar age gradient is shaped by the mass assembly history of the galaxy in the past, we interpret this as that $C_{\rm H\alpha}$ is a parameter not only tracing the star formation spatial distribution in the past $\sim 10$ Myrs as H$\alpha$ does, but likely also reflecting a more longer star formation pattern. For example, when a galaxy encounters a compaction event, it will maintain a high $C_{\rm H\alpha}$ for a quite long period. Thus we consider $C_{\rm H\alpha}$ as a useful indicator for studying the structure formation of galaxies.

At $z\sim0$, SFGs near the SFMS ridge line appear to have similar H$\alpha$ concentrations, with $C_{\rm H\alpha} \sim 0.4$, as shown in Figure~\ref{fig4}. This is because SFGs near the SFMS ridge are dominated by disk galaxies with $n\sim1$ \citep{Wuyts 2011}, thus having similar star formation spatial distributions. At the cosmic noon, there have been studies on the star formation spatial distribution \citep{Jain 2024}. At $z\sim0$, \citet{Wang 2019} have studied the star formation radial profiles of MaNGA SFGs with a stacking method. In the stellar mass range they probed ($M_{*}>10^{8.5}M_{\sun}$), SFGs near the SFMS ridge have similar SFR radial profiles. Our results are in line with those of \citet{Wang 2019}.

For central galaxies, the positive correlation between $C_{\rm H\alpha}$ and $n$ at low-mass regime is expected. As already well established, the cessation of star formation in the low-mass regime primarily acts on satellite galaxies via environmental effects, such as starvation and ram pressure stripping \citep{Peng 2010,Peng 2012}. In low-mass central galaxies, the stellar mass surface density ($\Sigma_{\ast}$) and star formation rate surface density ($\Sigma_{\rm SFR}$) follow a relatively tight positive correlation, i.e., the resolved star formation main sequence \citep{CanoDiaz 2016}. Thus the star formation distribution of a low-mass central galaxy is highly correlated with its morphology. In the high-mass regime, however, central galaxies begin to quench following an "inside-out" pattern \citep{Pan 2015,Belfiore 2017}, thus massive galaxies with high$-n$ do not necessary to have high star formation concentrations. A comparison between Figure~\ref{fig7} and Figure~\ref{fig4} suggests that many massive SFGs are on the path transforming from the cSF type into the cLIER type.

In the high-mass regime, we find disk-dominated SFGs with enhanced SFRs tend to have more centrally concentrated star formation distributions. This finding provides new insights in understanding the structure formation of massive galaxies. Using a sample of 394 MaNGA SFGs, \citet{Ellison 2018} investigated the stacked SFR radial profiles at different $\Delta_{\rm MS}$ bins. They found that SFGs with $\Delta_{\rm MS}>0.3$ have significant SFR enhancement at their central regions (see their figure 8), which is broadly consistent with the results shown in our Figure~\ref{fig4}. However, \citet{Ellison 2018} neither present the results for different mass bins, nor restrict to disk galaxies, perhaps due to their small sample size. \citet{Ellison 2018} interpreted their results under a "compaction" scenario, in which the central star formation enhancement is triggered via mergers or disk instability induced gas inflow \citep{Dekel 2014,Zolotov 2015,Tacchella 2016}. \citet{Wang 2019} have interpreted the significant central star formation variation of massive SFGs using the gas regulator model. They argued that this is because the gas depletion time is short in the central region of massive SFGs (see their figure 8). If this was the case, the central stellar mass buildup should be more efficient in massive SFGs. Regardless the detailed physics, Figure~\ref{fig8} and Figure~\ref{fig9} imply that compaction more frequently occurs in massive disk galaxies, eventually resulting a significant central mass buildup at log$(M_{\ast}/M_{\sun})>10.3$ as suggested in Figure~\ref{fig6}. This is also broadly consistent with the findings of \citet{Pan 2023}.
%Although our findings suggest that compaction events more frequently occur in the high-mass regime, the stellar mass growth during compaction events should be relatively small for $z\sim0$ massive disk SFGs. For a typical SFG with log$(M_{\ast}/M_{\odot})=10.5$ at $z\sim0$, its sSFR is $\sim$ log($\rm sSFR/yr^{-1})=-10.3$ \citep{Renzini 2015}. Assuming that the galaxy maintains its SFR, it takes 15 Gyrs to double its stellar mass. To significantly grow

Why there is no clear correlation between $C_{\rm H\alpha}$ and $\Delta_{\rm MS}$ at low-mass? We speculate low-mass galaxies are too efficient to build up disk components due to their rich \Hi~gas reservoirs. With a high \Hi~gas fraction, the disk build up time scale should be short in the low-mass regime. In observations, a high \Hi~gas fraction is coupled with a blue, extended disk component, supporting this picture \citep{Wang 2011, Chen 2020, Pan 2021}. When a low-mass galaxy moves to the upper envelope of the SFMS, the enhancement of SFR occurs across the entire galaxy, resulting in no significant changes in its SFR radial profile (see figure 5 of \citealt{Wang 2019}). Thus the enhancement of SFR leaves little impact on the structure of the galaxy. In observations, the majority of galaxies with log$(M_{\ast}/M_{\sun})<10.0$ have Sersic index $n<2.0$ (see Figure~\ref{fig4}), i.e., they are disk-dominated systems, suggesting that disk growth overwhelmingly dominates this mass regime. Following this picture, the more frequent compaction events exhibited in the high-mass regime is likely a natural consequence of inefficient disk growth compared to the low-mass regime, possibly due to the inefficient cool gas accretion. In the literature, the feedback from supermassive blackholes \citep{Croton 2006,Wang 2024} or halo shock heating \citep{Dekel 2006} are two most mentioned mechanisms for halting cool gas accretion in massive galaxies. More future observation and simulation works will shed light on this topic.

\section{Summary}
In this work, we present a study on the H$\alpha$ emission line flux concentration of $\sim 3000$ low$-z$ SFGs using the MaNGA IFS data. The H$\alpha$ flux concentration index $C_{\rm H\alpha}$ is determined using the H$\alpha$ flux growth curve. We have found the following.

1.~$C_{\rm H\alpha}$ is strongly correlated with the luminosity weighted stellar age gradient, supporting it as an useful indicator for tracing galaxy assembly history. SFGs near the star formation main sequence ridge line have the lowest $C_{\rm H\alpha}$, since they are dominated by pure disk systems with $n\sim1$.

2.~$C_{\rm H\alpha}$ is sensitive to environmental effects, in the sense that low-mass satellite galaxies below the star formation main sequence tend to have higher $C_{\rm H\alpha}$. This is in line with the "outside-in" star formation quenching mode exhibited in the low-mass regime as reported in the literature.

3.~Massive disk galaxies with enhanced star formation tend to have higher $C_{\rm H\alpha}$, while such a phenomenon is not seen in the low-mass regime. We interpret this as evidence that compaction events more frequently occur in the high-mass regime, which eventually resulting in the buildup of bulges in massive SFGs.

We conclude that $C_{\rm H\alpha}$ is a useful indicator for studying the ongoing central mass buildup of SFGs. Our findings provide new insights in understanding the structure formation picture of galaxies in different mass regimes.

\acknowledgments
%\begin{acknowledgments}
This work is supported by the National Key Research and Development Program of China (2023YFA1608100), the National Natural Science Foundation of China (NSFC, Nos. 12173088, 12233005 and 12073078).

Funding for the Sloan Digital Sky Survey IV has been provided by the Alfred P. Sloan Foundation, the U.S. Department of Energy Office of Science, and the Participating Institutions.SDSS-IV acknowledges support and resources from the Center for High Performance Computing  at the University of Utah. The SDSS website is www.sdss4.org. SDSS-IV is managed by the Astrophysical Research Consortium for the Participating Institutions of the SDSS Collaboration including the Brazilian Participation Group, the Carnegie Institution for Science, Carnegie Mellon University, Center for Astrophysics | Harvard \& Smithsonian, the Chilean Participation Group, the French Participation Group, Instituto de Astrof\'isica de Canarias, The Johns Hopkins University, Kavli Institute for the Physics and Mathematics of the Universe (IPMU) / University of Tokyo, the Korean Participation Group, Lawrence Berkeley National Laboratory, Leibniz Institut f\"ur Astrophysik Potsdam (AIP),  Max-Planck-Institut f\"ur Astronomie (MPIA Heidelberg), Max-Planck-Institut f\"ur Astrophysik (MPA Garching), Max-Planck-Institut f\"ur Extraterrestrische Physik (MPE), National Astronomical Observatories of China, New Mexico State University, New York University, University of Notre Dame, Observat\'ario Nacional / MCTI, The Ohio State University, Pennsylvania State University, Shanghai Astronomical Observatory, United Kingdom Participation Group,Universidad Nacional Aut\'onoma de M\'exico, University of Arizona, University of Colorado Boulder, University of Oxford, University of Portsmouth, University of Utah, University of Virginia, University of Washington, University of Wisconsin, Vanderbilt University, and Yale University.

%\end{acknowledgments}

\software{astropy \citep{astropy 2013,astropy 2018}, matplotlib  \citep{Hunter 2007} }

\appendix\label{sec:app}
In the main body of this work, we measure the concentration of H$\alpha$ flux based on the H$\alpha$ growth curve out to 1.5 $R_{\rm e}$. In this appendix we test whether our results are sensitive to the resolution effects. The angular FWHM of MaNGA sample is around $2.\arcsec 5$. Normalized by $R_{\rm e}$, this corresponds to $0.2-0.6$ $R_{\rm e}$ and $0.3-0.9$ $R_{\rm e}$ for the Primary and Secondary sample, respectively \citep{Bundy 2015}. In this work, we focus on the correlations among $C_{\rm H_{\alpha}}$, $M_{\ast}$ and $n$. Thus we investigate whether our measured $C_{\rm H_{\alpha}}$ show any additional dependence on resolution when $M_{\ast}$ and $n$ are fixed.

We first divide the cSF sample into two subsamples according to their morphologies, with $n<2.2$ and $n>2.2$. Figure~\ref{fig10} shows the $C_{\rm H_{\alpha}}-M_{\ast}$ relations for these two subsamples. We use the FoV parameter from the Pipe3D catalog to characterize the IFU resolution of a target galaxy. By definition, FoV is the IFU coverage normalized by $R_{\rm e}$. In the right column of Figure~\ref{fig10}, it can be seen that MaNGA galaxies have FoV larger than 1.5, i.e., the IFU coverage is out to at least 1.5 $R_{\rm e}$. Also, the FoV distribution shows two clear peaks, corresponding to the Primary and Secondary sample, respectively.

Symbols of the $C_{\rm H_{\alpha}}-M_{\ast}$ relation are color coded with FoV. By visually inspecting Figure~\ref{fig10} carefully, we find that at fixed $M_{\ast}$, the distribution of $C_{\rm H_{\alpha}}$ shows no clear dependence on FoV. This holds for galaxies with different Sersic index $n$. We thus conclude that the measurement of $C_{\rm H\alpha}$ is not sensitive to the resolution effect for the MaNGA sample.

\begin{figure*}
\centering
\includegraphics[width=140mm,angle=0]{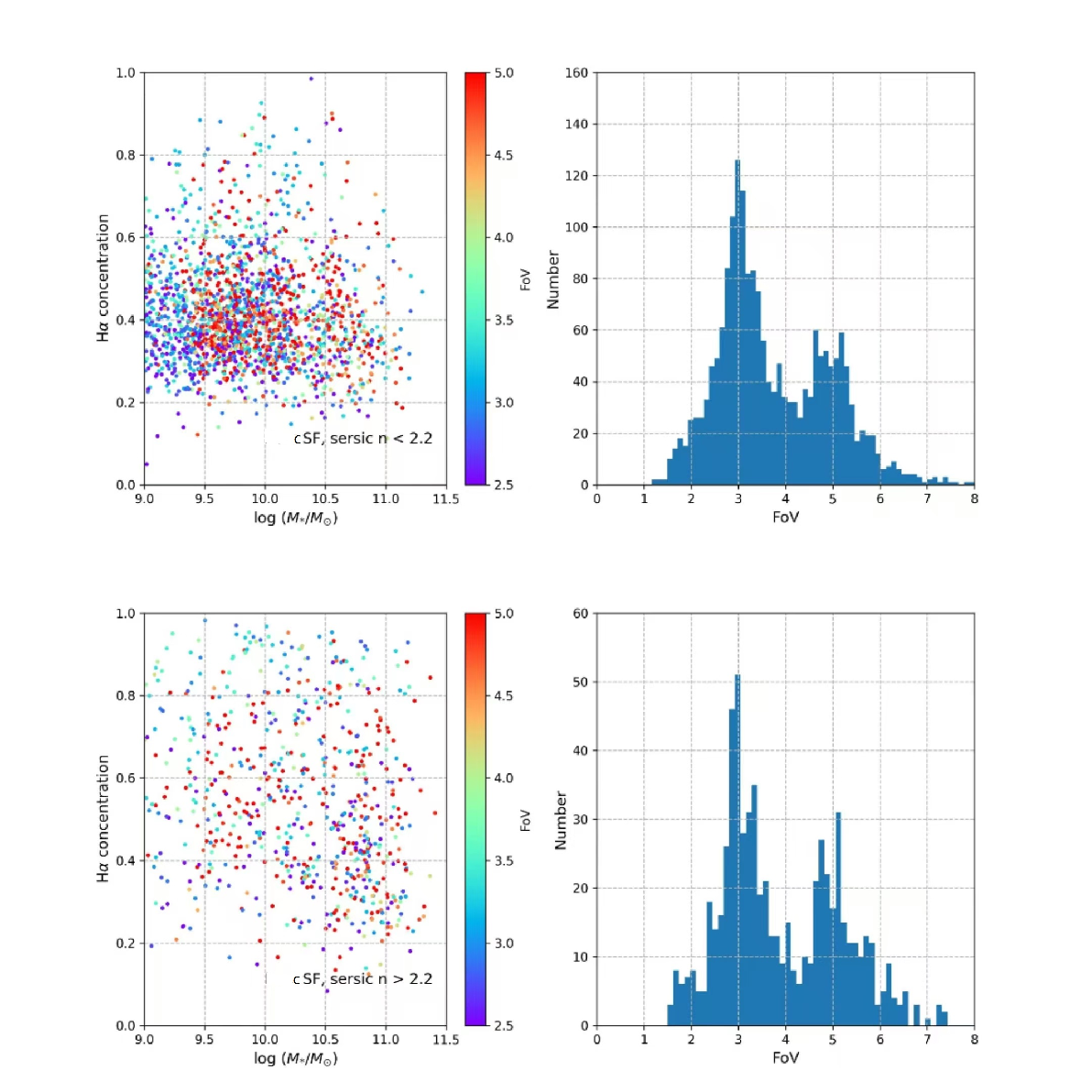}
\caption{We show the $C_{\rm H_{\alpha}}$ relation in two Sersic index bins ($n<2.2$ and $n>2.2$). The results are shown in the top left and top bottom panel. Symbols are color coded by the FoV parameter. The right column shows the FoV distribution.}
\end{figure*}\label{fig10}

\end{document}